# A Directed Signature Scheme and its Applications

Sunder Lal [*] and Manoj Kumar [**]

[*] *Dept of Mathematics, IBS Khandari.Dr. B.R.A.University Agra.*
Sunder_lal2@rediffmail.com.in.
[**] *Dept of Mathematics, HCST, Farah – Mathura,* (U. P.) – 281122.
Balyanyamu@rediffmail.com.in.

**Abstract -** This paper presents a directed signature scheme with the property that the signature can be verified only with the help of signer or signature receiver. We also propose its applications to share verification of signatures and to threshold cryptosystems.

## 1. Introduction

To avoid forgery and ensure confidentially of message, it has been a common practice for centuries for the sender of the message to put his identification marks, *i.e.* signature on the letter and then to seal it in an envelope, before handing it over to a deliverer. In electronic-era, physical signatures are not workable and the **digital signatures** are the cryptographic answer to the problems of information security and authenticity. W.Diffie and M.Hellman originally defined digital signatures. Digital signatures are in some sense similar to hands written signature. A hand written signature is verified by comparing it to others authentic signatures. In contrast to hand written signatures, which are independent of message, digital signature must somehow reflect both the message and the signer.

In order to achieve this requirement, a signer who holds secret information, generates digital signature, using secret information. To verify digital signature the receiver applies a publicly known algorithm. Thus, a signature scheme allows a user with a public key and a corresponding private key to sign a document in such a way that everyone using public key can verify the signature of the document, but no one else can forge the signature. This property is called as self-authentication, which is necessarily required for some applications of digital signatures such as certificate official announcements issued by some authority.

On the other hands, there are so many situations, when the signed message may be sensitive to the signature receiver. Signatures on medical records, tax information and most personal/business transactions are such situations. Such signatures are called as **directed signatures** [1, 2, 9, 10, and 12]. In directed signature scheme, the signature receiver has full control over the signature verification process. Nobody can check the validity of signature without his cooperation.

The concept of directed signatures was first presented by C.H.Lim and P.J. Lee [10]. It is a construction based on the GQ signature scheme [8]. D.Chaum [2] introduced the concept of designated confirmation. Later T.Okamoto [12] presented a more practical construction of designated confirmer signatures.

This paper proposes **a directed signature scheme and its applications**. In this scheme, any third party can check the signature validity with the help of signature receiver or the signer as well. Both the signer and signature receiver have full control over the signature verification process. In other words, they are independent to prove the validity of the signature to any third party, whenever necessary. The paper is organized as follows: -

The section-2 presents **some basic tools**. Section-3 describes a **Directed Signature Scheme.** The **security** of the proposed scheme is discussed in section-4**.** Section – 5 is an illustration. **A directed signature with threshold verification** is discussed in section - 6. An **application to threshold cryptosystem** is discussed in section-7.

## 2. Preliminaries

*2.1. Preliminary Settings*

Throughout this paper we use the following system setting.

- A prime modulous $p$, where $2^{511} < p < 2^{512}$;
- A prime modulous $q$, where $2^{159} < q < 2^{160}$ and q is a divisor of $p – 1$;
- A number $g$, where $g \equiv k^{(p-1)/q} \mod p$, $k$ is random integer with $1 \leq k \leq p-1$ such that $g > 1$; (g is a generator of order $q$ in $Zp^*$).
- A collision free one-way hash function $h$ [15];

The parameters $p, q, g$ and $h$ are common to all users. We assume that every user A chooses a random $x_A \in Zq$ and computes $y_A = g^{x_A} \mod p$. Here $x_A$ is the private key of A and $y_A$ is the public key of A. For our purpose, we use Schnorr's signature scheme [13] and Shamir's threshold scheme [14]. These basic tools are briefly described below.

*2.2. Schnorr's signature scheme*

In this scheme, the signature of A on message *m* are given by ($r_A, S_A$),

where, $r_A = h(g^{k_A} \mod p, m)$ and $S_A = k_A - x_A \cdot r_A \mod p$.

Here random $k_A \in Zq$ is private to A. The signature are verified by checking the equality

$$r_A = h(g^{S_A} y^{r_A} \mod p, m).$$

*2.3. Shamir's threshold scheme*

Shamir's (*t, n*) threshold secret sharing signature scheme is a scheme to distribute a secret key K into *n* users in such a way that any *t* users can cooperate to reconstruct K but a collusion of *t* – 1 or less users reveal nothing about the secret. Shamir's scheme is based on Lagrange interpolation in a field. To implement it, a polynomial *f* of degree *t* – 1 is randomly chosen in $Z_q$ such that *f* (0) = K. Each user *i* is given a public identity $u_i$ and a secret share $f(u_i)$. Now any *t* out of *n* shareholders can reconstruct the secret K = *f* (0), by pooling their shares and using

$$f(0) = \sum_{i=1}^{t} f(u_i) \prod_{j=1, j\neq i}^{t} \frac{-u_j}{u_i - u_j} \mod q$$

Here for simplicity the authorized subset of *t* users consists of shareholders *i* for *i* =1,2,3…t.

## 3. New directed signature scheme

Suppose that user A wants to generate a signature on message *m* so that only receiver B can directed verify the signature and that B as well as A can independently prove the validity of signature to any third party C, whenever necessary. Our construction of such a signature scheme is based on Schnorr's signature scheme. The signing and verification processes are as follows.

*3.1. Signature generation by A*

(a). A picks random $K_{a_1}$ and $K_{a_2} \in Z_q$ and computes

$$W_B = g^{-K_{a_2}} \mod p \text{ and } V_B = g^{K_{a_1}} \cdot y_B^{K_{a_2}} \mod p.$$

Here $y_B$ is the public key of the signature receiver B.

(b). Using a one-way hash function *h*, A computes a secret value $r_A = h(g^{K_{a_1}}, m)$.

(c). A computes $S_A = K_{a_1} + x_A \cdot r_A \mod q$.

Here $x_A$ is the private key of the signer. { $S_A$, $W_B$, $V_B$, *m*} is the signature of A on the message *m*.

*3.2. Signature verification by B*

(a). Using his private key $x_B$, B computes $R = V_B (W_B)^{x_B} \mod p$ and recovers $r_A = h(R, m)$.

(b). B checks the following congruence for a valid signature $g^{S_A} \stackrel{?}{\equiv} R \cdot y_A^{r_A} \mod p$.

If hold then $\{S_A, W_B, V_B, m\}$ is a valid signature.

### 3.3. Proof of validity to C

In this scheme, both the signer and signature receiver can independently prove the validity of signature to any third party, whenever necessary. This sub-section describes the protocol using which the signer and the signature receiver can prove the validity of signature.

#### 3.3.1. Proof of validity by A to C

1. A computes $V_C = g^{K_{a_1}} \cdot y_C^{K_{a_2}} \mod p$ and sends to C.

2. C uses $V_C$ in place $V_B$ to checks the validity of signature by using his secret key. The signature verification process will remain same as in sub-section 3.2.

#### 3.3.2. Proof of validity by B to C

1. B picks random $K \in Z_q$ and computes

$$W_C = g^{-K} \mod p, \quad V_C = R \cdot y_C^K \mod p,$$ and sends to C.

2. C uses $(W_C, V_C)$ in place $(W_B, V_B)$ to checks the validity of signature by using his secret key. The signature verification process will remain same as in sub-section 3.2.

## 4. Security discussions

In this sub-section, we shall discuss the security of proposed **Directed Signature Scheme.**

**(a).** Can one retrieve the secret key $x_A$, integer $K_{a_1}$ from the equation

$$S_A = K_{a_1} + x_A \cdot r_A \mod q \,?$$

Here the number of unknown parameters is two. The number of equation is one, so it is computationally infeasible for a forger to collect the secret $x_A$, integer $K_1$ from this equation. Obviously, this is also again computationally infeasible for a forger to collect any information.

**(b).** Can one impersonate the signer?

A forger may try to impersonate the signer by randomly selecting two integers $K_1$ and $K_2 \in Z_q$ and calculate

$$W_B = g^{-K_2} \mod p, \quad V_B = g^{K_1} \cdot y_B^{K_2} \mod p, \quad r_A = h(g^{K_1}, m).$$

But without knowing the secret key $x_A$, it is difficult to generate a valid $S_A$ to satisfy the verification equation $g^{S_A} \stackrel{?}{\equiv} [R \cdot y_A^{r_A}] \mod p$.

**(c).** Can one forge a signature $\{S_A, W_B, V_B, m\}$ by the equation $g^{S_A} \stackrel{?}{\equiv} [R \cdot y_A^{r_A}] \bmod p$?

A forger may randomly select an integer R and then computes the hash value $r_A$ such that

$$r_A = h(R, m) \bmod q.$$

Obviously, to compute the integer $S_A$ is equivalent to solving the discrete logarithm problem. On the other hand, the forger can randomly select $r_A$ and $S_A$ first, then try to determine a value $R^*$, that satisfy the signature verification equation.

Thus these attacks will not be successful.

## 5. Illustration

We choose smaller parameters to illustrate the scheme. Taking $p = 23$, $q = 11$ and $g = 3$. The secret and public key of users is as follow----

| User | Secret key | public key |
|------|------------|------------|
| A    | 4          | 12         |
| B    | 7          | 2          |
| C    | 6          | 16         |

*5.1. Signature generation by A to B*

(a). A picks random $K_{a_1} = 9$, and $K_{a_2} = 5$ and computes $W_B = 16$, $V_B = 1$.

(b). Using a one way hash function $h$, A computes $r_A = 10$, $S_A = 5$.

(c). A sends $\{5, 16, 1, m\}$ to B as his/her signature on the message $m$.

*5.2. Signature verification by B*

(a). B computes $R = 18$, recovers $r_A = h(18, m) = 10$.

(b). B checks the following congruence for a valid signature $3^5 \stackrel{?}{\equiv} 18 \cdot 12^{10} \bmod 23$. This holds.

*5.3. Proof of validity by A to C*

1. A computes $V_C = 16$ and sends to C.

2. C computes $R = 16 \cdot 16^6 \bmod 23 = 18$.

3. C uses $V_C$ in place $V_B$ to checks the validity of signature by using his secret key. The signature verification process will remain same as in sub-section 3.2.

*5.4. Proof of validity by B to C*

1. B picks random $K = 8$ and computes $W_C = 4$, $V_C = 9$, and sends to C.

2. C computes $R = 9 \cdot 4^6 \mod 23 = 18$.

3. C uses $(W_C, V_C)$ in place $(W_B, V_B)$ to checks the validity of signature by using his secret key. The signature verification process will remain same as in sub-section 3.2.

## 7. Application to shared verification of signatures

The proposed directed signature scheme can be used to design a signature scheme with threshold verification. A signer can generate a signature on the message $m$ to the organization $R$ in such a way that any subset $H_R$ of $k$ users of a designated group $G_R$ of $n$ users can determine the validity of the signature but any $k-1$ or less users cannot verify the signature The construction of **directed signature scheme with threshold verification** is based on Shamir's threshold scheme [14]. We assume that every user $i$ in the designated group $G_R$ possesses a private key $x_{R_i}$ and public key $y_{R_i}$. This scheme consists of the following steps.

*7.1. Signature generation by A*

(a). A picks random $K_{a_1}$ and $K_{a_2} \in Z_q$ and computes

$$W_R = g^{-K_{a_2}} \mod p \text{ and } V_R = g^{K_{a_1}} \mod p.$$

(b). Using a one-way hash function $h$, A computes a secret value

$$r_A = h(V_R, m) \text{ and } S_A = K_{a_1} + x_A \cdot r_A \mod q.$$

Here $x_A$ is the private key of the signer.

(c). A selects a polynomial $f_R(x) = K_{a_1} + b_1 x + \ldots b_{k-1} x^{k-1} \mod q$, with $K_{a_1} = f_R(0)$.

(d). A computes a public value $v_{R_i}$ for each member of the group $G_R$, as,

$$v_{R_i} = f_R(u_{R_i}) \cdot y_{R_i}^{K_{a_2}} \mod p.$$

Here $y_{R_i}$ is public key and $u_{R_i}$ is the public value associated with each user $i$ in the group $G_R$.

(e). A sends $\{S_A, W_R, m, \{v_{R_i}\}_{i=1}^{n}\}$ to the group $G_R$ as the signature of A on the message $m$.

*7.2. Signature verification by the organization R*

Any subset $H_R$ of $k$ users from a designated group $G_R$ of $n$ users can verify the signature. We assume that there is a designated combiner, which collects partial computations from each user in $H_R$ and determine the validity of signature. The verifying process is as follows.

(a) Each user $i \in H_R$ recovers his/her secret share, as, $f_R(u_{R_i}) = v_{R_i} W_R^{x_{R_i}} \mod p$.

(b) Each user $i \in H_R$ modifies his/her shadow, as, $MS_{R_i} = f_R(u_{R_i}) \cdot \prod_{j=1, j \neq i}^{t} \frac{-u_{R_i}}{u_{R_i} - u_{R_j}} \mod q$.

(c) Each user $i \in H_R$ uses his/her modified shadow, $MS_{R_i}$, and calculate the partial result $R_{R_i}$, as,

$$R_{R_i} = g^{MS_{R_i}} \mod q,$$

and sends to the combiner.

(d). The combiner computes $R$, as, $R = \prod_{i=1}^{k} R_{R_i} \mod p$ and recovers $r_A$, as,

$$r_A = h(R, m) \mod q.$$

(e). The combiner checks the following congruence for a valid signature

$$g^{S_A} \stackrel{?}{\equiv} R \cdot y_A^{r_A} \mod p.$$

If hold then $\{S_A, W_R, m, \{v_{R_i}\}_{i=1}^{n}\}$ is a valid signature.

*7.3. Remarks*

The proposed scheme has the following characteristics:

- There is no need of a trusted SDC, while in many threshold signature schemes, there is a trusted SDC that determines the security parameters and also distributes the secrets shares to each user in the system.
- The signer is free to decide the security parameters and secrets shares to each user.
- The signer decides the security parameters and secrets shares at the time of signing.
- The signer can use the security parameters and secrets shares as one-time secrets. There is no need to fix them.
- The signer can change the threshold value $k$ to sign another documents.
- The only requirement is that every user $i$ in the designated group $G_R$ possesses a private and public key pair $(x_{R_i}, y_{R_i})$.

## 8. Application to threshold cryptosystem

Threshold cryptosystem is society – oriented cryptosystem. This is useful, when a policy requires the consent of more than one person. The organization has a single public key. The secret key is distributed among the users of organization in such a way that any $k$ users can determine the policy, but when any $k-1$ users work together, they cannot receive any information about the system secret. Hence any $k-1$ users cannot determines the policy.

Desmedt. Y. and Frankel.Y. [4, 5, 6] developed the concepts of threshold cryptosystems. They're some weaknesses in these cryptosystem, given as follows:

- The group secret key is fixed; any subgroup of *k* dishonest users can recovers the group secret key and can be harmful for the system.
- The security parameters are also fixed. The secrets shares determine during the system set up. Any change in-group member or security policy requires adjusting group member's secret share accordingly.

The proposed directed signature scheme can be used to make a threshold cryptosystem to sort out all the above weaknesses.

Suppose a user A wants to encrypt the message *m* so that any *k* users from designated group $G_R$ of *n* users should cooperate to recover the message. The encryption and decryption key *K* for sender and the receiving group $G_R$ can be recovered from $V_R$ and *R* respectively. The sender A broadcasts {{ $S_A$, $W_R$, *c*, $\{v_{R_i}\}_{i=1}^{n}$ } to the group $G_R$, where $c = E_K(m)$ and $K = h(V_R)$. Thus, any *k* members can recover *R* and then decryption key $K = h(R)$. They can decrypt the message as $m = D_K(c)$.

The proposed threshold cryptosystem has the following advantages over the cryptosystems based on the concept of Desmedt and Frankel.

- The group secret is not fixed and can be change for each further communications.
- There is no need of fixed set up. The sender determines everything at the time of encryption.
- An individual can send cipher text to any chosen group with any desired security policy.

## Reference

1. Boyar, J., Chaum D., Damgard I. and Pederson T., (1991), Convertible undeniable signatures. *Advances in Cryptology – Crypto,* 90*,* LNCS # 537,p.p.189-205.

2. Chaum D. (1995). Designated confirmer signatures, *Advances in Cryptology Euro crypt, 94* LNCS # 950,p.p.86-91.

3. Chaum D. (1991). Zero- knowledge undeniable signatures. *Advances in Cryptology – Eurocrypt, 90,* LNCS # 473,p.p.458-464.

4. Desmedt, Y. and Frankel Y. (1991). Shared Generation of Authenticators and Signatures. In *Advances in Cryptology –Crypto -91, Proceedings.* p.p. 457-469. New York: Springer Verlag.

5. Desmedt, Y. (1988). Society and group oriented cryptography. In *Advances in Cryptology – Crypto -87, Proceedings.* p.p. 457-469. New York: Springer Verlag.

6. Desmedt, Y. (1994). Threshold cryptography. *European Transactions on Telecommunications and Related Technologies.*Vol. 5,No. 4, p.p.35 – 43.

7. Diffie W. and Hellman M. (1976), New directions in Cryptography, *IEEE Trans. Info.Theory.*31.pp. 644 - 654.

8. Guillou, L.C. and Quisquater J.J. (1988), A practical zero-knowledge protocol fitted to security microprocessors minimizing both transmission and memory. *"Advances in Cryptology –Eurocrypt, 88,* LNCS # 330,p.p.123 - 128.
9. Lim C.H. and Lee P.J. (1993). Modified Maurer-Yacobi, scheme and its applications. *Advance in cryptology –Auscrypt,* LNCS # 718, p.p. 308 – 323.
10. Lim C.H. and P.J.Lee. (1996). Security Protocol, In Proceedings of International Workshop, (Cambridge, United Kingdom), Springer-Verlag, LNCS # 1189.
11. Mullin R.C., Blake I.F., Fuji – Hara R. and Vanstone S.A. (1985). Computing Logarithms in a finite field of characteristic two. *SIAM J. Alg.Disc.Meth.,* p.p.276 – 285.
12. Okamoto T. (1994), Designated confirmer signatures and public key encryption are equivalent. *Advances in Cryptology – Crypto, 94* LNCS # 839, p.p.61-74.
13. Schnorr C.P. (1994). Efficient signature generation by smart cards, *Journal of Cryptology,* 4(3), p.p.161-174.
14. Shamir A. (1979). How to share a secret, *communications of the ACM*, 22: p.p. 612 - 613.
15. Zheng, Y., Matsummoto T. and Imai H. (1990). Structural properties of one – way hash functions. *Advances in Cryptology – Crypto, 90,* Proceedings, p.p. 285 – 302, Springer Verlag.

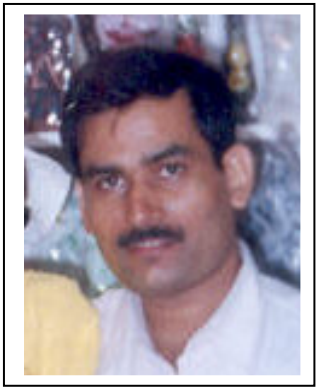

**Manoj Kumar** received the B.Sc. degree in mathematics from Meerut University Meerut, in 1993; the M. Sc. in Mathematics (Goldmedalist) from C.C.S.University Meerut, in 1995; the M.Phil. (Goldmedalist) in *Cryptography*, from Dr. B.R.A. University Agra, in 1996; submitted the Ph.D. thesis in *Cryptography*, in 2003. He also taught applied Mathematics at DAV College, Muzaffarnagar, India from Sep, 1999 to March, 2001; at S.D. College of Engineering & Technology, Muzaffarnagar, and U.P., India from March, 2001 to Nov, 2001; at Hindustan College of Science & Technology, Farah, Mathura, continue since Nov, 2001. He also qualified the *National Eligibility Test* (NET), conducted by *Council of Scientific and Industrial Research* (CSIR), New Delhi- India, in 2000. He is a member of Indian Mathematical Society, Indian Society of Mathematics and Mathematical Science, Ramanujan Mathematical society, and Cryptography Research Society of India. His current research interests include Cryptography, Numerical analysis, Pure and Applied Mathematics.